# Four Factor Authentication with emerging cybersecurity for Mobile Transactions

**Sanyam Jain[1], Raju gautam[2], Shivani Sharma[3], and Ravi Tomar[4]**

   **Abstract**— Cybersecurity is very essential for Mobile Transactions to complete seamlessly. Mobile Commerce (Mcom.) is the very basic transaction type, which is very commonly used (2 in 5 people uses mobile as transaction medium), To secure this there are various technologies used by this research. The four "factors" formally known as Multi-Factor-Authentication are: two of them are Traditional methods (User Login-password and One Time Password (aka OTP) with addition of Geolocation and Facial Recognition. All the data is converted to a text file, which is hidden in an image (using Babushka algorithm). The end-point then decrypts the image using same algorithm.

## 1. Introduction

THIS research paper intends to make mobile transactions more secure using multifactor authentication. Utilizing mobile resources (sensors) optimally at time of Transaction.

The whole module (Program) will run on the Bank's gateway server. The current technology we use in 2018 (2-FA) can be extended with these two more factors namely Geolocation and Facial recognition. The old methods are very easy to break. Once a hacker (Thief) gets your sim everything is leaked. According to an experiment (Research by Wired), Once a sim is compromised (so as sms) one can get your Google. There are various methods then to retrieve your text message for e.g. PushBullet is used to monitor your mobile activity to computer to increase mobility. There are various proved Methods to crack sim Text Messages using intermediary networks like "Man in the Middle attack". Thus, two more methods need to be increased to make transactions more secure there is geolocation matcher and facial recognition matcher, which altogether matches the user current location of point of purchase as well as the face, which is stored in bank database.

## 2. Till Now

   There are various techniques used by banking institutions to secure mobile transactions. Banks usually comes with two basic solutions called 2FA includes user details (generally user ID password and another is one time password) for end customer/user. These details are generally breakable at very low-level computation. End user on the bank website sets user name and password and the 4-6 digit code is received on phone registered with bank. Till now there is no additional security for this. Every mobile transaction is directly or indirectly depends on only mobile phone. This strategy worked well when there were people with no smartphones. The user

Affiliation: School of Computer Science, University of Petroleum and Energy Studies, Dehradun, India
Email:[1]sanyamjaincs@gmail.com,[2]rajugautam45@gmail.com,
[3]1998shivisharma@gmail.com, [4]ravitomar7@gmail.com

detail gets encrypted and transferred to bank in secured environment. There are high rates of compromising and cracking bank transactions. Day by day banks updates their security algorithms even then also the transactions being compromised either by networks or by phishing user authenticity credentials.

## 3. Proposed Work

To overcome these low level security measures, in this paper two advanced methods are proposed so as to wipe hacking of mobile transactions. User after performing basic two factors i.e. user credentials and OTP it'll get redirected to the gateway consisting of geolocation (Performed background) and Facial Recognition in main stream. All will be done on the smartphone/ smart hand held device consisting of GPS (To match location) and Front Camera (For facial recognition). The Computation part will run on Bank server, the customer will only get the instance of program. Starting from second factor, The geolocation part, user will have to update the current location of performing transaction to the bank database which is personalized for each user of bank – a dashboard. The same location is matched with the bank server at time of purchasing by the customer. This will run in background and user will not even get notified about the location factor. (This is done to make transaction faster and less informative so as user thinks of three factors).All the data gets converted to a text file. Text file consists of Four things : Username + Password + Image ASCII code(Fetched from facial Recognition) + GeoLocation (Of the mobile phone) The text file then embedded/encrypted with the image file and sent to the Merchant. Merchant or Receiving bank then Decrypts the Image file, gets the user information, and authenticate that. This way the transaction is completed and successfully transfers money. The actual program should deploy on the server, which executes several instances for each invoking user, and same methodology occurs every time.The proposed work is different among all the existing work by "Babushka/Matryoschka" algorithm. The algorithm has been developed under many trainings to get text file encoded to image file in the meta data of the image. The encryption uses HMAC-SHA256 to authenticate the hidden data.

## 4. Methodology

This section of the paper consists of working of all four factors described each one by one as following:

1.  User Login and Password ➔ this is very basic authentication factor used to verify the user against the database. The local database or internet database anything can work to store and verify the user credentials. (Here we used sqlite3 for Ubuntu) The username and password are matched with database and are verified and proceeded to next step.

2.  Mobile Authentication or One Time Password ➔ OTP helps the bank to authenticate user accessibility to the point of sale as to determine the ownership or privacy for the original user to continue to transaction. OTP may be 4 or 6 digit code as per standard. OTP is now submitted to the portal. OTP is hosted by the Twilio API (here). There are numerous services for the same. OTP is just requested to the Twilio server and user buffers a hash key for same. Now since hash can only be broken by Twilio service, Thus hash and the OTP received are matched on the device

3. Facial Recognition ➔ The Third Milestone is the Facial recognition system as the authentication factor.
   After authenticating OTP the user will be asked for his/her image capture, the camera automatically takes it and stores it in the file system for the future authentication purpose. Now the webcam analyzes the registered users face in the organization database through the webcam and identifies his features, thereby unlocking the third milestone as secured user. Facial recognition is used along with OpenCV library as the source tool for identifying the Human's face.(Link: https://opencv.org )This technique which allows very less chance to the attacker to bypass a particular system.

4. Geolocation (Of the mobile phone) ➔ the geolocation part, user will have to update the current location of performing transaction to the bank database which is personalized for each user of bank – a dashboard. The same location is matched with the bank server at time of purchasing by the customer. This will run in background and user will not even get notified about the location factor. This creates the integrity of the location from which a user operates. The smartphone app (here : Location Tracker for androiod) will be in sync to the organization or bank database server which will be operating on hosted website. The User manually after the OTP validation will update the Location Tracker app with the Update button on it, this will set the latitude, the longitude of the user's current location based on the GPS, and this will be reflected in the database server. This is a crucial step for a user whenever attacker tries to use VPN as a layer to infiltrate into the user account, The security will be hard to crack as mobile device GPS is ON and this matches the transaction done from the application used to perform transaction.

## 4.1. Existing Security

The traditional approach towards doing transaction or logging in is being entering username and strong alphanumeric passwords. but they are vulnerable to different attacks like social engineering, rainbow table , brute force attack etc. This approach also have OTP as a second factor sometimes but they are liable to Man in the Middle attack which causes leaking of the transferred OTP. There is a need to replace this technique with some additional liabilities.Devoted programmers have little issue bypassing through the weaker executions, either by capturing codes or by misusing account-recuperation frameworks. The guarantee of two-factor started to disentangle right off the bat. By 2014, lawbreakers focusing on Bitcoin administrations were discovering courses around the additional security, either by catching programming tokens or more detailed record recuperation plans. Now and again, assailants followed telephone bearer accounts specifically, setting up a minute ago call-sending game plans to capture codes in travel.

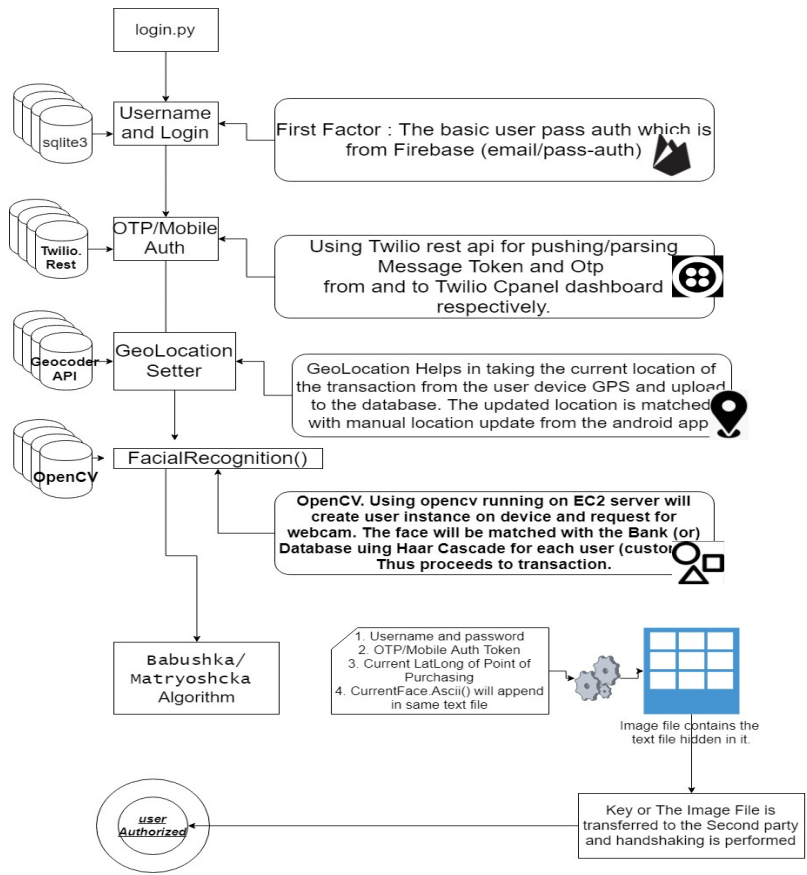

*Figure 1. The working model*

## 4.2.  Proposed Security

4FA aims at higher assurance in comparison to single-factor schemes. Our proposed system uses multiple factor which acts as a upper model in the shortcomings of alphanumeric and OTP factors. The 4FA will be using facial Recognition and The geolocation feature as extra milestone in addition to the above mentioned technique. Geolocation creates the integrity of the location from which the user operates.

1. If OTP is verified, system asks for face recognition and   run geolocation check in background automatically.
2.  If OTP is not verified, system stops transaction.
3.  If the face recognition fails, system stops transaction.
4. The user verifies location through app, if geolocation verifies, then proceed the user, else system stops transaction.

The utilization of different verification methods to demonstrate one's personality depends on the commence that an unapproved on-screen character is probably not going to have the capacity to supply the components required for. On the off chance that, in a confirmation endeavor, no less than one of the parts is missing or provided incorrectly, the client's personality isn't built up with adequate sureness being ensured by multi-factor validation at that point stays blocked.

## 4.3.  Matryoschka Algorithm

At last the application will collect all data in a text file and the text file is sent to the merchant for successful transaction. The intermediary is "Babushka/Matryoschka" algorithm.Matryoschka are basically sets of traditional Russian wooden dolls of decreasing size placed one inside the other. A Matryoschka doll can be opened to reveal a smaller figure of the same sort inside, which has, in turn, another figure inside, and so on.The Russian Matryoschka Museum recently exhibited a collection of similarly designed Matryoschka sets, differing only in the number of nested dolls in each set. Unfortunately, some over-zealous (and obviously unsupervised) children separated these sets, placing all the individual dolls in a row. There are n dolls in the row, each with an integer size. You need to reassemble the Matryoschka sets, knowing neither the number of sets nor the number of dolls in each set. You know only that every complete set consists of dolls with consecutive sizes from one to some number m, which may vary between the different sets. Matryoschka algorithm invokes Encryption HMAC-SHA256 bit to authenticate the hidden data which further implements the XTEA algorithm. To hide the sensitive data. All the previous authentication factors will be passed through this encryption tunnel in order to hide information in the image form. The text will not be visible throughout the transfer of image, due to the hard coded encryption by HMAC-SHA256 and XTEA algorithm. The non-significant values of the pixels on image are used to store all sensitive information being encrypted by XTEA algorithm which is being sent to the sever for completing the transaction. The SHA in turn is being implemented to check the integrity of the data sent.

    1. The Username and password from first factor are captured.

    2. The data from second factor will be i.e., OTP will be selected.

    3. The data from third factor i.e., facial recognition, the image captures will be used, but first its conversed to ASCII ART of their image.

    4. The data from fourth factor i.e. latitude and longitude will be captured.

All the above mentioned three points will be first stored in a text .txt file as the primary source and then this algorithm automatically encrypts that text data inside the selected image,  so as to preserves its identity and then it is being send to merchant. All the above-mentioned four points will be first stored in a text .txt file as the primary source and then this algorithm automatically encrypts that text data inside the selected image, to preserves its identity.

    **Syntax: python babushka.py –hide –m <mac-password> -k <password> <secret> <change>**

But first organization have to install python library called pillow.py using  "sudo pip install pillow", this will install all the dependencies required for Matryoschka. The file then gets encrypted and stored, into a mentioned image, as a factor for high security. Now, let's say if a merchant wants to decode that image there is module called DecodeNow.py, which will only be used by merchant for the pure decoding purpose

    **Syntax: python babushka.py –open –m <mac-password> -k <password> <image>**

## 4.4. Software Engine Specification

1. The software described in this application is based on the python client integrated with third party messaging service API i.e., twilio.

2. OTP is just requested to the Twilio server and user buffers a hash key for same and is secured for sending OTP messages.

3. Python being more popular language can be integrated in any system making four factor authentication more scalable.

4. The API use makes it easy for system to communicate with other services. One of the API being used is Geolocation API of google.

5. The geolocation API uses different techniques to identify user's current location like public IP address of the device, cell tower IDs, GPS information, a list of Wi-Fi access points, signal strengths and MAC IDs (Wi-Fi and/or Bluetooth).

6. The open source android technology is used for the geolocation factor. The app location tracker  of our institution will be in sync to the organization/bank database server which will be operating on hosted website

7. The machine leaning technique is being implemented on the system for the face recognition factor through a image dataset , which in turn are trained through a variety of positive and negative images, thereby generating a common standard format i.e. XML which contains facial feature (Cascade). This format can be used with any technology thereby making system scalable and flexible.

8. The XML recognizes a real time face. And Frame of the dataset can be stored as for further use.

9. PyQt4 is used to give the graphical Dialog Box Layout to the system performing transactions. Though it can be modified as per Application's need because of Open Source development.

## 5. Result

The system is developed with integration of different modules. The complete transaction costs at end user is 30 seconds. And complexity at bank server costs *O(n)* .

## 6. Conclusion

The use of multi-factor authentication helped to achieve the purpose of more secure electronic transactions. The four-factor validation arrangement prepares client by giving solid validation to expansive scale. The 4FA could unquestionably decrease the recurrence of online fraud and data leakage furthermore, other online blackmails in the light of the way that the causality secret key could never again be adequate to give a criminal access to user information .